\documentclass[3p,times,twocolumn]{elsarticle}

\usepackage{ecrc}

\volume{00}

\firstpage{1}

\journalname{Nuclear and Particle Physics Proceedings}

\runauth{}

\jid{nppp}

\jnltitlelogo{Nuclear and Particle Physics Proceedings}

\usepackage{amssymb}

 \biboptions{comma,square,sort&compress}

\usepackage[figuresright]{rotating}

\begin{document}

\begin{frontmatter}



\dochead{}

\title{Extracting $\hat{q}$ from single inclusive data at RHIC and  at the LHC for different centralities: a new puzzle?}


\author[label1]{Carlota Andr\'es}
\author[label1]{N\'estor Armesto}
\author[label1,label2]{Matthew Luzum}
\author[label1]{Carlos A. Salgado}
\author[label1,label3]{P\'ia Zurita}

\address[label1]{Instituto Galego de F\'\i sica de Altas Enerx\'\i as IGFAE, Universidade de Santiago de Compostela, E-15782 Santiago de Compostela (Spain)}
\address[label2]{Instituto de F\'\i sica - Universidade de S\~ao Paulo, Rua do Mat\~ao Travessa R,
no. 187, 05508-090, Cidade Universit\'aria, S\~ao Paulo (Brasil)}
\address[label3]{Physics Department, Brookhaven National Laboratory, Bldg. 510A, Upton, NY 11973, USA}

\begin{abstract}
We present here an extraction of the jet transport coefficient, $\hat{q}$, using RHIC and LHC single-inclusive high-$p_T$ data for different centralities. We fit a $K$-factor that determines the deviation of this coefficient from an ideal estimate, $K \equiv \hat{q}/(2\epsilon^{3/4})$, where $\epsilon$ is given by hydrodynamic simulations. As obtained already in previous studies, this $K$-factor is found to be larger at RHIC than at the LHC. However it is, unexpectedly, basically no-dependent on the centrality of the collision. Taken at face value this result, the $K$-factor would not depend on local properties of the QGP as temperature, but on global collision variables such as the center of mass energy.
\end{abstract}


\end{frontmatter}


\section{Introduction}
Jet quenching has been established as a successful tool to extract medium parameters that describe the QGP created in heavy-ion collisions. This talk is based on our recent publication \cite{Andres:2016iys}, where we perform an extraction of the $\hat{q}$ parameter using RHIC and LHC data on the nuclear modification factor, $R_{\rm AA}$, for single-inclusive particle production at high transverse momentum. The formalism of Quenching Weights \cite{Baier:2001yt,Salgado:2002cd,Salgado:2003gb}, embedded in different hydrodynamic descriptions of the medium, is used.\\

We define the jet quenching parameter $K \equiv \hat{q}/(2\epsilon^{3/4})$, motivated by the ideal estimate $\hat q_{\rm ideal}\sim 2\epsilon^{3/4}$ \cite{Baier:2002tc}, where $\epsilon$ is the energy density given by the (three) hydrodynamic models . Our main results are that this $K$-factor is $\sim2 - 3$ times larger for RHIC than for the LHC and, surprisingly, it is almost independent of the centrality of the collision.

\section{Energy loss implementation}
\label{sect:energyloss}
The production of a hadron $h$ at transverse momentum $p_T$ and rapidity $y$ can be described by
\begin{eqnarray}
\frac{d\sigma^{AA\rightarrow h+X}}{dp_Tdy} &=& \int \frac{dx_2}{x_2}\frac{dz}{z}\sum\limits_{i,j}x_1f_{i/A}(x_1,Q^2)x_2f_{j/A}(x_2,Q^2)\times\nonumber\\
&\times& \frac{d\hat{\sigma}^{ij\rightarrow k}}{d\hat{t}}D_{k \rightarrow h}(z, \mu_F^2)\ .
\label{crosssection}
\end{eqnarray}

All the calculations are done at NLO using the code \cite{Stratmann:2001pb}, with the proton PDF set CTEQ6.6M \cite{Nadolsky:2008zw} and EPS09 nuclear correction \cite{,Eskola:2009uj}. The renormalization, fragmentation and factorization scales are taken as $\mu_F = p_T$.The energy loss is absorbed in a redefinition of the fragmentation functions: 
\begin{eqnarray}
D^{(med)}_{k\rightarrow h}(z,\mu_F^2)=\int \limits_0^1d\epsilon P_E(\epsilon)\frac{1}{1 - \epsilon}D^{(vac)}_{k\rightarrow h}\left(\frac{z}{1 - \epsilon},\mu_F^2\right)\ ,
\label{medfragmentations}
\end{eqnarray}
where $P_E(\epsilon)$ are the ASW Quenching Weights and $D^{(vac)}_{k\rightarrow h}(z,\mu_F^2)$, DSS vacuum fragmentation functions \cite{deFlorian:2007aj}.\\
The ASW Quenching Weights, i.e. the probability distribution of a fractional
energy loss, $\epsilon = \Delta E/E$, of the fast parton in the medium, are based on two main assumptions: gluon emissions are independent and fragmentation functions are not modified. It has been found that for the total coherence case and for soft radiation these are good approximations \cite{CasalderreySolana:2012ef,Blaizot:2012fh,Armesto:2007dt}. In fact, for soft radiation and no finite energy effects QW and rate equations are equivalent. In our analysis, the QW are used in the multiple soft or harmonic oscillator approximation.\\

The quenching weights, $P_i(\Delta E/\omega_c,R)$  are tabulated in \cite{Salgado:2003gb} for the case of a static medium of finite length $L$ and transport coefficient $\hat q$, where
\begin{equation}
\omega_c=\frac{1}{2}\hat qL^2, \quad  R=\omega_cL\ .
\label{eq:omcR}
\end{equation}

For a expanding medium, if $\hat q(\tau)\sim 1/\tau^\alpha$, a dynamic scaling law was found \cite{Salgado:2002cd} that relates the resulting spectra with an equivalent static scenario. Hence, $\omega_c^{eff}$ and $R^{eff}$ for a hydrodynamic medium profile are computed as
\begin{eqnarray}
\omega_c^{eff}(x_0,y_0,\tau_{\rm prod},\phi)=\int d\xi\,\xi\,\hat q(\xi),\\
R^{eff}(x_0,y_0,\tau_{\rm prod},\phi)=\frac{3}{2}\int d\xi\,\xi^2\, \hat q(\xi).
\label{eq:omceff}
\end{eqnarray}

So, we only need to specify the relation between the local value of the transport coefficient $\hat q(\xi)$ at a given point of the trajectory and  the hydrodynamic properties of the medium:

\begin{equation}
\hat q(\xi)=K\cdot 2\epsilon^{3/4}(\xi),
\label{eq:qhatlocal}
\end{equation}
where $K\simeq 1$ would correspond to the ideal QGP \cite{Baier:2002tc}. The local energy density $\epsilon(\xi)$ is taken from a hydrodynamic model of the medium, for which we will consider several different options in the next sections. However, there is an ambiguity on its value for times smaller than the thermalization time $\tau_0$ . Consequently, we consider here three different extrapolations for the time from the hard production to the thermalization time:
\\
\noindent (i) $\hat q(\xi)=0$ for $\xi<\tau_0$;
\\
\noindent (ii) $\hat q(\xi)=\hat q(\tau_0)$ for $\xi<\tau_0$; and
\\
\noindent (iii) $\hat q(\xi)=\hat q(\tau_0)/\xi^{3/4}$ for $\xi<\tau_0$, usually called the free-streaming extrapolation.
\\

\section{Hydrodynamic models of the medium}
\label{sect:hydro}
We obtain the space-time distribution of the local energy density by solving the relativistic hydrodynamic equations. To check that our conclusions do not depend on the hydrodynamic profile used, three different hydrodynamic setups are employed.

The first, which we refer to as ``Hirano'', corresponds to the calculation described in \cite{Armesto:2009zi,Hirano:2001eu,Hirano:2002ds,hydro-site}.  In short, this computation uses an optical Glauber model with initial proper time $\tau_0 = 0.6$ fm and with vanishing viscosity.

The other two hydrodynamic models correspond to the calculations in \cite{Luzum:2008cw}. Both of them start at an initial proper time of $\tau_0 = 1$ fm and use an equation of state inspired by lattice QCD calculations. One simulation, which we refer to as ``Glauber'', uses for an initial condition an energy density proportional to the density of binary collisions, while the ratio of shear viscosity to entropy density is fixed to a constant value of $\eta/s = 0.08$. The final computation is referred to as ``fKLN''. This  takes its initial condition from a factorised Kharzeev-Levin-Nardi model, with the shear viscosity set to $\eta/s = 0.16$.

\begin{figure}
\includegraphics[width=0.5\textwidth]{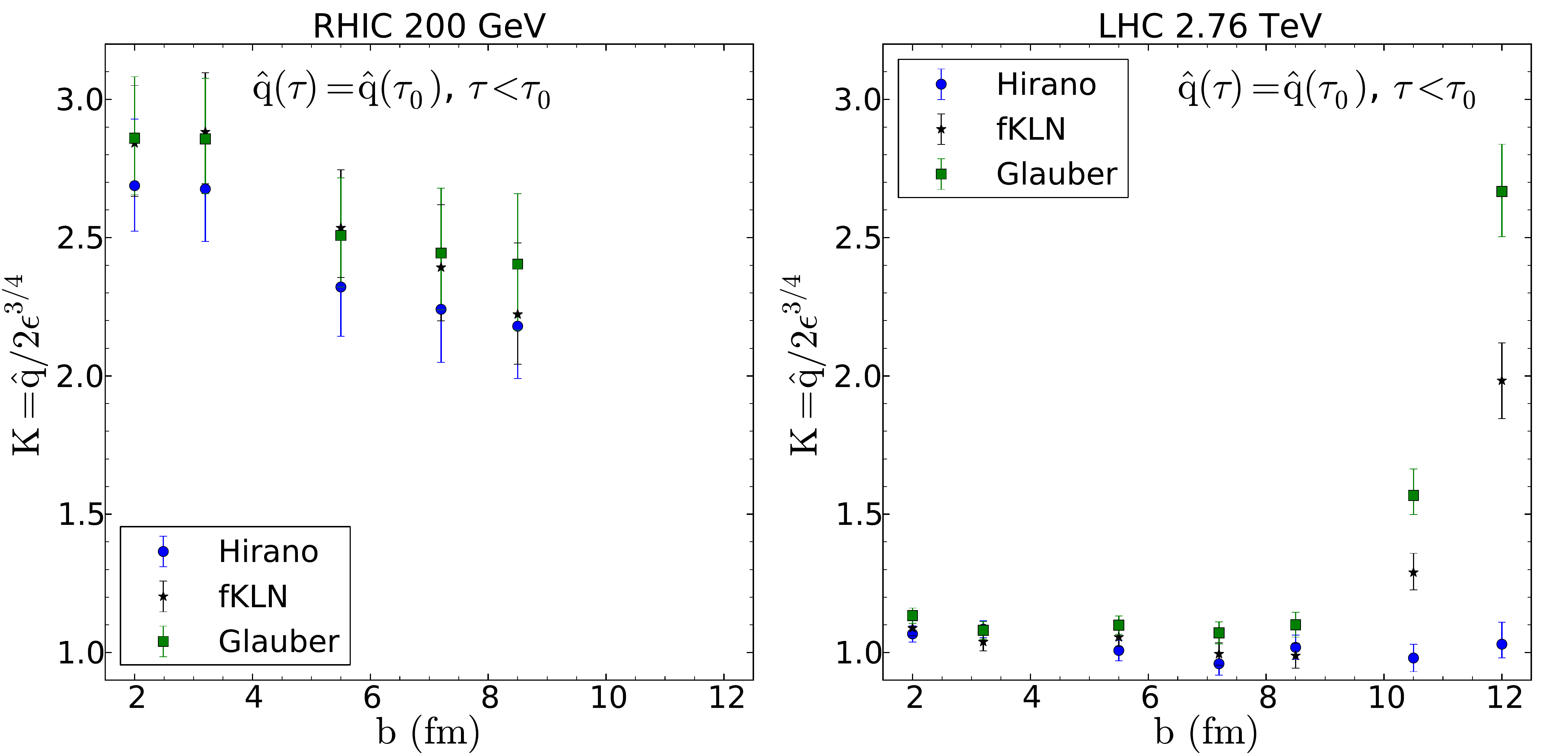}
\caption{$K$-factors obtained from fits to PHENIX $R_{\rm AA}$ data \cite{Adare:2008qa} \textit{(left panel)} and to ALICE $R_{\rm AA}$ data \cite{Abelev:2012hxa} \textit{(right panel)} using different hydrodynamic profiles vs. the average impact parameter for each centrality class and $\hat{q}$ constant before thermalization, see the previous Sections.}
\label{fig:chi2RHICandLHC}
\end{figure}

\section{Results}
\label{sect:results}

We restrict our analysis to the simplest observable $R_{\rm AA}$ at RHIC \cite{Adare:2008qa} and the LHC \cite{Abelev:2012hxa}. It is the first study of the centrality dependence of both LHC and RHIC data.

\begin{figure}
\includegraphics[width=0.5\textwidth]{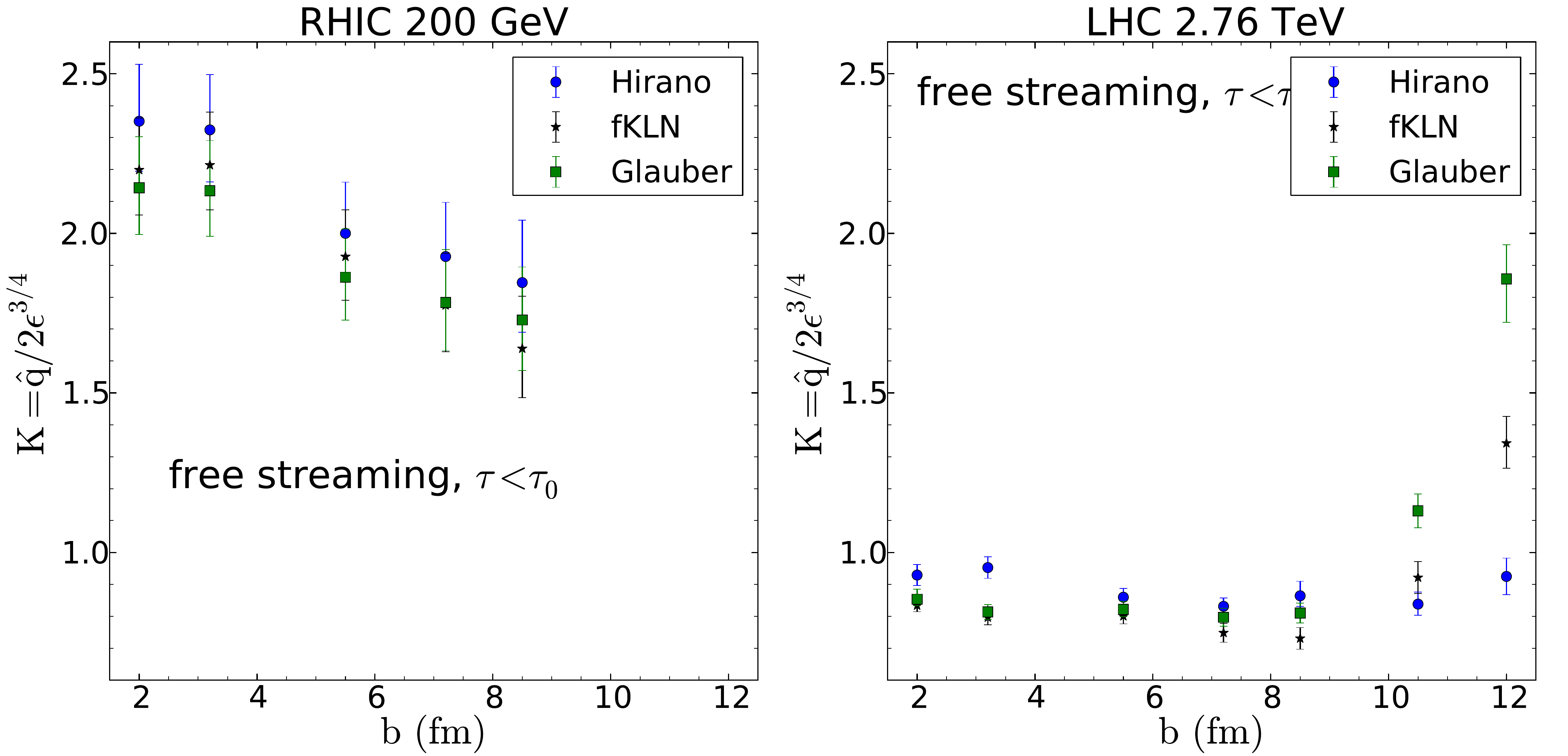}
\caption{$K$-factors obtained from fits to PHENIX $R_{\rm AA}$ data \cite{Adare:2008qa} \textit{(left panel)} and to ALICE $R_{\rm AA}$ data \cite{Abelev:2012hxa} \textit{(right panel)} using different hydrodynamic profiles vs. the average impact parameter for each centrality class and for the free-streaming extrapolation, see the previous Sections.}
\label{fig:chi2RHICandLHCfreestreaming}
\end{figure}

\begin{figure}
\includegraphics[width=0.5\textwidth]{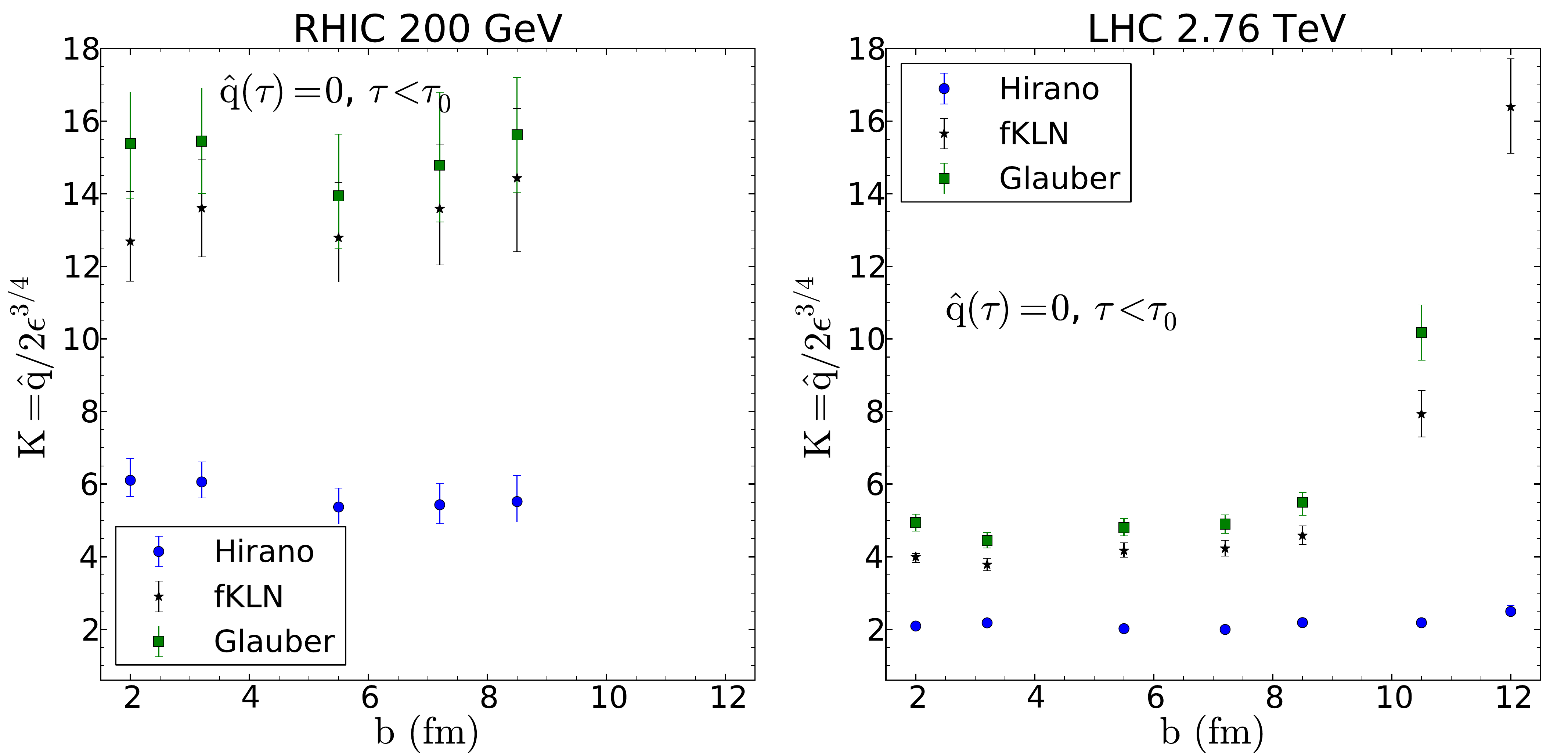}
\caption{$K$-factors obtained from fits to PHENIX $R_{\rm AA}$ data \cite{Adare:2008qa} \textit{(left panel)} and to ALICE $R_{\rm AA}$ data \cite{Abelev:2012hxa} \textit{(right panel)} using different hydrodynamic profiles as a function of the average impact parameter for each centrality class and for $\hat q(\xi)=0$ before thermalization, see the previous Sections.}
\label{fig:chi2RHICandLHCqhat0}
\end{figure}

We have performed a $\chi^2$ fit to the best value of $K$ for each energy and centrality using three different hydrodynamic profiles and three different assumptions for the time prior to the equilibration time, see the previous Sections. The uncertainty band is determined by $\Delta \chi ^2 = 1$. In the left panels of Fig.~\ref{fig:chi2RHICandLHC}, Fig.~\ref{fig:chi2RHICandLHCfreestreaming} and Fig.~\ref{fig:chi2RHICandLHCqhat0} we plot the different values of the K-parameter fitted to the PHENIX data \cite{Adare:2008qa} for different combinations of hydrodynamic profiles and behavior before the thermalization time. The corresponding values for the LHC \cite{Abelev:2012hxa} are plotted in the right panels of the same figures.

First of all, the values of $K$ obtained are compatible for Fig.~\ref{fig:chi2RHICandLHC} and Fig.~\ref{fig:chi2RHICandLHCfreestreaming}, and the results for the three different hydrodynamic setups are similar. Second, the results at RHIC are flat or slightly decreasing with decreasing centrality, while at the LHC the behavior is constant except for the most peripheral collisions where it depends very much on the hydrodynamic model employed. Third, the $K$-factor is $\sim2 - 3$ times larger for RHIC than for the LHC. Other groups \cite{Burke:2013yra} have found a factor $\sim 25\%$. Therefore, the extracted value of $K$ seems to depend mainly on the energy of the collision and to be independent of the centrality of the collision.

\begin{figure*}
\includegraphics[width=\textwidth]{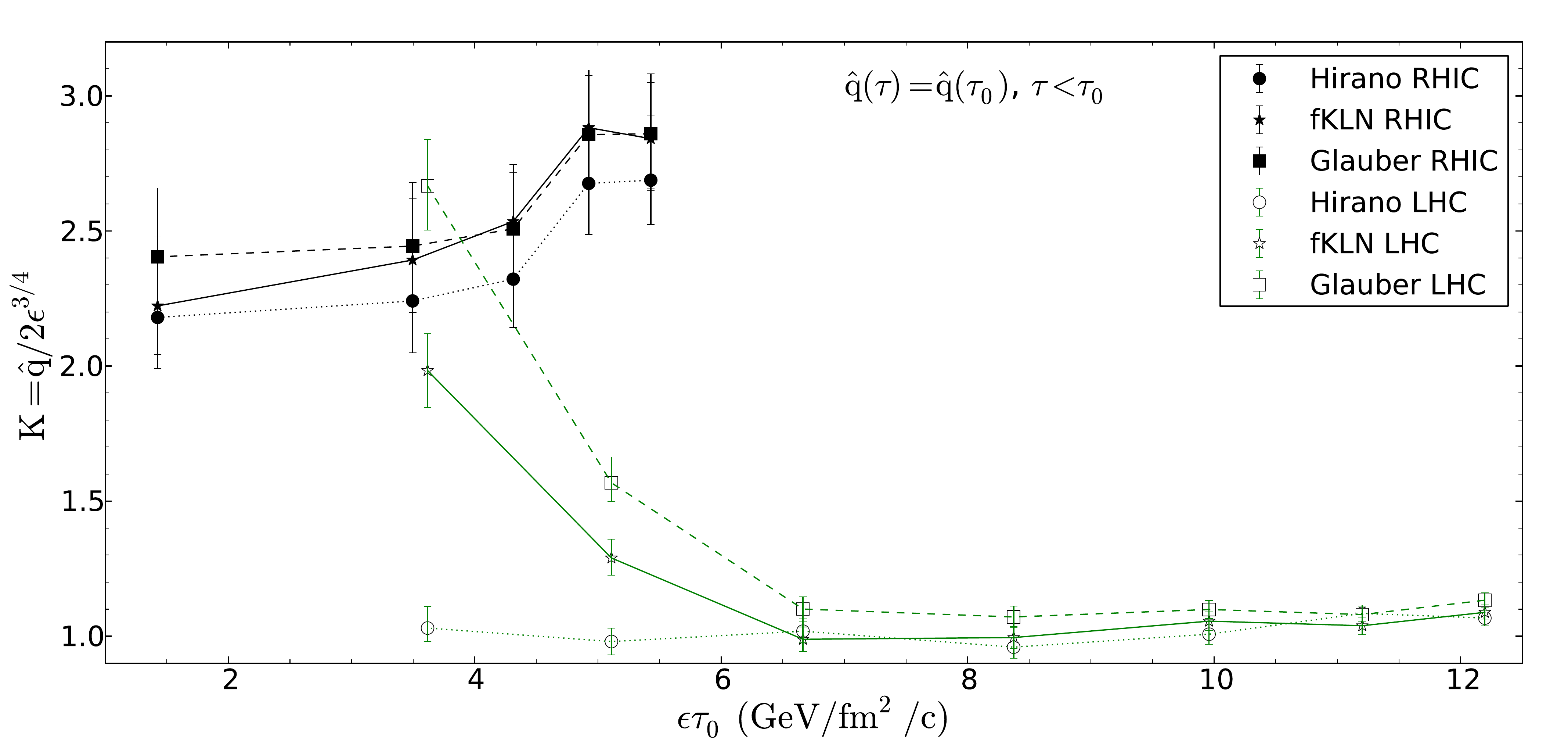}
\caption{$K$-factor at RHIC and LHC energies for different centrality classes versus an estimate of the energy density \cite{Adare:2015bua,Adam:2016thv} times formation time $\tau_0$ of the QGP created in each case.}
\label{fig:overlapRHIC-LHC}
\end{figure*}

As there is an overlap on typical energy densities (or temperatures) between central AuAu at RHIC and semi-peripheral PbPb at the LHC, in a naive interpretation, their values of $K$ should coincide. To illustrate that this interpretation does not correspond to the present findings we plot in Figure \ref{fig:overlapRHIC-LHC}, the $K$-factors obtained for different centralities and energies versus an energy density times formation time $\tau_0$ extracted from the experimental data using Bjorken estimates \cite{Adare:2015bua,Adam:2016thv}.

\section{Conclusions}
\label{sect:conclusions}
We have analyzed the single-inclusive suppression of particles at high transverse momentum as a function of centrality and the energy of the collision. A constant $K$-factor with respect to the perturbative estimate $\hat q\simeq 2\epsilon^{3/4}$ is defined. This factor is fitted to the corresponding experimental data at RHIC and LHC for different centralities. The  obtained value at the LHC is close to unity, while the one at RHIC confirms large corrections to the ideal case. 

As the medium formed at the LHC has a larger temperature, one may be tempted to naively interpret that it is closer to the ideal case than the one at RHIC, for which larger corrections, could be needed. Nevertheless, the centrality dependences at RHIC and the LHC separately are rather flat, that is, the change in the value of $K$ is not only due to the different temperature, as there is a large region of overlap between RHIC and the LHC for different centralities. 

Our approach has several limitations that may affect the results. First of all, as we have already mentioned, the quenching weights are based on two assumptions which could fail if color coherence is broken. They are computed in the multiple soft scattering approximation, where the perturbative tails of the distributions are neglected, which may enhance the energy loss. The scaling law used has only been proved for $\hat{q}(\tau) \propto 1/\tau$. Collisional energy loss is also neglected in our formalism.

The CMS collaboration presented new experimental data on $R_{ \rm AA}$ of PbPb collisions at $\sqrt{s} = 5.02$ TeV at InitialStages2016 in Lisbon. These data show that the suppression at the LHC does not depend on the center of mass energy of the collision. To extract the exact value of $K$ from these data, a new fit needs to be performed. However, we have checked that this $K$-factor would be $\sim10\%$ smaller than the corresponding one at $\sqrt{s} = 2.76$ TeV.

\section*{Acknowledgements}
This research was supported by the European Research Council grant HotLHC ERC-2011-StG-279579; the People Programme (Marie Curie Actions) of the European Union's Seventh Framework Programme FP7/2007-2013/ under REA grant agreement \#318921 (NA); Ministerio de Ciencia e Innovaci\'on of Spain under project FPA2014-58293-C2-1-P; Xunta de Galicia (Conseller\'{\i}a de Educaci\'on) --- the group is part of the Strategic Unit AGRUP2015/11. C. Andr\'es thanks the Spanish Ministery of Education, Culture and Sports for financial support (grant FPU2013-03558).


\nocite{*}

\bibliographystyle{elsarticle-num}
\bibliography{jos}

\end{document}